\documentclass[preprint,showpacs,preprintnumbers,amsmath,amssymb,nofootinbib,showkeys,aps]{revtex4}
\usepackage{graphicx}% Include figure files
\usepackage{dcolumn}% Align table columns on decimal point
\usepackage{bm}% bold math
\usepackage{latexsym}
\usepackage{epsfig}
\usepackage{amsmath}
\usepackage{color}
\usepackage{hyperref}
\usepackage{mathrsfs}
\newcommand{\be}{\begin{equation}}
\newcommand{\ee}{\end{equation}}
\newcommand{\like}{\mathscr{L}}
\newcommand{\texpdf}{\texorpdfstring}

\begin{document}

\title{Cosmological Constraints on Scalar Field Dark Matter}

\author{A. A. Escobal$^{1}$}\email{anderson.aescobal@gmail.com}
\author{J. F. Jesus$^{1,2}$}\email{jf.jesus@unesp.br}
\author{S. H. Pereira$^{1}$}\email{s.pereira@unesp.br}

\affiliation{$^1$Universidade Estadual Paulista (UNESP), Faculdade de Engenharia, Guaratinguet\'a, Departamento de F\'isica e Qu\'imica - Av. Dr. Ariberto Pereira da Cunha 333, 12516-410, Guaratinguet\'a, SP, Brazil
\\$^2$Universidade Estadual Paulista (UNESP), Campus Experimental de Itapeva - R. Geraldo Alckmin 519, 18409-010, Itapeva, SP, Brazil,
}

\begin{abstract}
%\begin{center}{\bf ABSTRACT}\end{center}
This paper aims to put constraints on the parameters of the Scalar Field Dark Matter (SFDM) model, when dark matter is described by a free real scalar field filling the whole Universe, plus a cosmological constant term. By using a compilation of 51 $H(z)$ data and 1048 Supernovae data from Panteon, a lower limit for the mass of the scalar field was obtained, $m \geq 5.1\times 10^{-34} $eV and $H_0=69.5^{+2.0}_{-2.1}\text{ km s}^{-1}\text{Mpc}^{-1}$. Also, the present dark matter density parameter was obtained as $\Omega_\phi = 0.230^{+0.033}_{-0.031}$ at $2\sigma$ confidence level. The results are in good agreement to standard model of cosmology, showing that SFDM model is viable in describing the dark matter content of the universe.
\end{abstract}

\maketitle

\section{Introduction}

Astronomical observations provide us with important informations about the evolution of the universe. The type Ia Supernovae (SNe Ia) data endow us with strong evidences of the current accelerated expansion of the universe due to dark energy (DE) and the spectrum of Cosmic Microwave Background (CMB) radiation states that the large-scale universe is homogeneous and isotropic at least for anisotropies up to order of $10^{-5}$. Using the Friedmann-Robertson-Walker metric (FRW) and the General Relativity equations, we obtain the standard cosmological model, named $ \Lambda$CDM model, which describes an universe nearly spatially flat and currently undergoing an accelerating phase due to recent domination of a component with negative pressure represented by $\Lambda$. This component represents DE, which corresponds to about $70\%$ of the energy content of the Universe, while Cold Dark Matter (CDM) in the halo of galaxies represents $ 25 \% $ and the ordinary baryonic matter corresponds to about $ 5 \% $. Currently, the contribution of radiation is negligible.

This is the model that, in general, satisfactorily describes observational data, however it presents some problems due to misunderstanding of the nature of DM \cite{p8, pp} and DE \cite{W1, W2}. This stimulates the search for new models describing these unknown components.

Bosonic scalar fields are important objects of study in several branches of theoretical physics, with applications from Quantum Field Theory to Cosmology. They are present since the origin of the Universe, for instance the fundamental scalar fields describing the grand unified theory (GUT) \cite{GL, LCF} and, soon after, in the evolution of the universe, being responsible for inflation and reheating  mechanisms \cite{GA, LA}. In addition, some authors have also assumed the nature of DM and DE as essentially represented by scalar fields \cite{me1, me2, me3, me4, me5}. Although only the Higgs boson and some mesons have been detected experimentally in nature as true scalar fields, other scalar fields could be present in the universe, as predicted by supersymmetric models.

Scalar Field Dark Matter models, which treats the DM as formed by a real scalar field $ \phi $ minimally coupled to gravity subject to a potential $ V (\phi) $, have provided interesting results recently \cite{1,2, js, 3}. In \cite{1,2} it has been proposed that DM is an ultralight zero spin boson with a Compton wavelength of the order of few kpc, condensing at high temperatures of the order of $\sim$ TeV, leading to formation of Bose-Einstein condensates in the early universe. Thus, droplets of DM were formed, which would become the halos of galaxies \cite{hg}. This implies all galaxies to be very similar, having their halos formed practically at the same time at high redshift. The DM scalar field mass estimated was of about $10^{-22}$ eV. A more recent work \cite{t1} has found an inferior limit for the SFDM mass of the order of $10^{-32}$ eV, a value close to the lower limit found in \cite {js} of $m > 10^{-33}$ eV, using combination of SNe Ia \cite{s1} and $H(z)$ \cite {h1} data.

%Thus, it has been found that the rotation profile of the halos of DM \cite{1,2} are compatible with those observed. Hydrodynamic analysis of scalar field density fluctuations was also made using a given potential $ V (\phi) $, with a mass of $ 10-^{22}$ eV for DM particle. In a linear perturbation regime, the SFDM model reproduces the same results of the $ \Lambda$CDM model, differing only in non-linear regimes. In these cases, the SFDM model has a flat density profile while $ \Lambda$CDM has higher formation of galaxies and clusters of galaxies in the center.

In this work we study the recent evolution of the universe driven by a SFDM model in which the DM particle is represented by a real, massive and homogeneous scalar field $ \phi $, which is minimally coupled to gravity. In section \ref{dynamics}, we begin by introducing the dynamic equations of the SFDM model. In section \ref{approx}, we develop an approximation in order to obtain more efficient numerical solutions for the field. In section \ref{data}, the observational data used is briefly described. In section \ref{analysis}, the model is analysed according to Bayesian Statistics \cite{b1, b2}, using the Monte Carlo Markov Chain (MCMC) method \cite{M1, M2, M3, M4} as a tool. Finally, in section \ref{conclusion}, we present the conclusions.

% iiiiiiiiiiiiiiiiiiiiii  rever o ultimo paragrafo

%We performed the individual and combined analysis of the H(z) data of sample \cite{Hz} and the over 1000 data of SNs Ia \ref{} to obtain a lower bound for DM mass , as seen in the sections (??). To make the numerical resolution more efficient when analyzing large values of DM mass, in which the numerical analysis becomes slow due to the oscillations of the field, we propose in section III a JWKB approximation, where we can find an expression for the field $ \phi $. Are we in the section ?? that the SFDM model is a good alternative to $ \Lambda$CDM, since it is able to solve some problems of $ \Lambda$CDM, in addition to reproducing the good results of $ \Lambda$CDM. All we have to do is hope that in the future the astronomical observations will reveal the origin of DM.

\section{Dynamics of SFDM\label{dynamics}}

The dynamic equations of the SFDM model plus baryonic ordinary matter and a cosmological constant $\Lambda$ term are, in a spatially flat FRW background \cite{js}:
\begin{align}
\dot H&=-\frac{\kappa^2}{2}\left(\dot{\phi}^2+\rho_{b}\right),\label{back1}\\
%\label{DC}
\ddot{\phi} + 3H \dot{\phi} + \frac{dV}{d\phi}&=0,\label{back2}\\
{\dot\rho_{b}}+ 3H \rho_{b}&=0, \label{back3}\\
{\dot\rho_{\Lambda}}&=0, \label{back4}
\end{align}
together with Friedmann constraint:
 \begin{equation}
H^2=\frac{\kappa^2}{3}\left(\rho_{\phi}+\rho_{b} + \rho_{\Lambda} \right)\, ,
\label{eq:FC}
\end{equation}
where $\rho_b$, $\rho_{\phi}$ and $\rho_{\Lambda}$ correspond to the energy densities of baryonic matter, scalar field and cosmological constant term, with $\kappa^{2} \equiv 8\pi G$ and $H \equiv \dot{a}/a$ the Hubble parameter.

Using a quadratic potential of the form $V(\phi)=m^{2}\phi^{2}/2$,
where $m$ is the physical mass of the scalar field, and using the change of variables \cite{js},
\begin{eqnarray}
r\cos \theta \equiv \frac{\kappa}{\sqrt{6}}\frac{\dot{\phi}}{H},\,\,\,r\sin \theta \equiv \frac{\kappa}{\sqrt{6}}\frac{m\phi}{H},\,\,\,\theta \equiv \tan^{-1}\frac{m \phi}{\dot{\phi}}, \,\,\,
b \equiv \frac{\kappa}{\sqrt{3}}\frac{\sqrt{\rho_{b}}}{H},\,\,\,
l \equiv \frac{\kappa}{\sqrt{3}}\frac{\sqrt{\rho_{\Lambda}}}{H},
\label{eq:varb}
\end{eqnarray}
we can rewrite the dynamic equations of the model, (\ref{back1})-(\ref{back4}), as: 
\begin{align}
r'&= -3r\cos^2(\theta) +\frac{3}{2}\Pi r, \label{eq:pol_r} %\label{EQSN1}
\\
\theta'&= s +\frac{3}{2}\sin(2\theta), \label{eq:pol_q}\\
b'&=\frac{3}{2}\left(\Pi -1\right)\,b, \label{eq:pol_b}\\
l'&=\frac{3}{2}\Pi l, \label{eq:pol_l}\\
s'&={3\over 2}\Pi s. \label{eq:pol_s}%\label{EQSN2}
\end{align}
where $\Pi=2r^2\cos^2(\theta)+b^2$ and $ s $ is a new parameter, defined by $s\equiv m/H$. A prime denotes derivative with respect to $ N \equiv \ln a $, the ``e-folding number".

Finally, we complete the new system of equations, (\ref{eq:pol_r})-(\ref{eq:pol_s}), with a sixth equation that allows us to analyse SNe Ia data, which depends on the luminosity distance $D_L$ given in terms of the comoving distance, $D_C$:
\begin{equation}
D_L=(1+z)\frac{H_0}{c}D_{C}\,,
\end{equation}
where $c$ is the speed of light. Since $ D_C$ depends on $ H (z) $ but we do not have an analytic expression for $ H (z) $, we need a differential equation for $ D_C$. For a spatially flat universe we can write:
\begin{equation}
\label{clark}
\frac{dD_{C}}{dz}\equiv \frac{1}{E(z)},
\end{equation}
with $E(z)\equiv\frac{H(z)}{H_0}=\frac{\mu}{s}$ and $\mu=m/H_0$. Regarding the independent variable $ N $, we have:
\begin{equation}
 D'(N)=-\frac{e^{-N}s}{\mu} \,,
\end{equation}
which corresponds to the sixth equation for the dynamic system (\ref{eq:pol_r})-(\ref{eq:pol_s}). The initial conditions at $ N = 0 $ (today) are given by the vector of parameters $\vec{r}_0$:
\begin{equation}
\vec{r}_0=\left(\sqrt{\Omega_{\phi0}},\theta_0,\sqrt{\Omega_{b0}},\sqrt{1-\Omega_{\phi0}-\Omega_{b0}},\mu\right) \,.
\end{equation}

As well known \cite{js}, this leads to a high oscillatory behaviour of the field, a behaviour which corresponds to an average equation of state (EOS) corresponding to dust, $\langle\omega\rangle\approx0$. In order to make the numerical solution of the system \eqref{eq:varb} more efficient, next we seek for an approximation in the mass interval that we are interested, namely, $m\gg H_0$ ($\mu\gg1$).

\section{JWKB approximation\label{approx}}

Assuming a time dependent $ \eta $ function, $ \eta(t) $, we can write the scalar field dynamic equation (\ref{back2}) as:
\begin{eqnarray}
    \label{JWKB}
    \phi''(\eta) +\phi'(\eta)\left( \frac{\ddot{\eta}+3H \dot{\eta}}{\dot{\eta}^2} \right) + \frac{m^2 \phi}{\dot{\eta}^2} &=& 0 \, ,
\end{eqnarray}
where we use $\phi'(\eta) \equiv \frac{d\phi}{d\eta}$. In order to use the JWKB approximation we must eliminate the term $ \phi'(\eta)$, so we must cancel the term in parentheses of the above relation. This term vanishes when
\be
\ddot{\eta}+3H \dot{\eta} = 0,\nonumber
\ee
whose solution is
\be
    \dot{\eta}= Ca^{-3}.
\ee
where $C$ is a constant. Thus, equation (\ref{JWKB}) becomes:
\begin{eqnarray}
    \phi''(\eta) + \frac{m^2 a^6 \phi}{C^2}=0 \, .
\end{eqnarray}

The above equation is of the form,
\begin{eqnarray}
    \frac{d^2y(x)}{dx^2} + h^2 \psi(x)y(x) = 0 \,,
\end{eqnarray}
with $h=m/C$ a constant. The approximate solution is given by \cite{Jeffreys25,Ratra91}:
\begin{eqnarray}
    y(x) &=& \psi(x)^{-\frac{1}{4}}(A\cos{L}+B\sin{L})\, ,\\
    L &=& \int_0^x \psi(x)^{\frac{1}{2}}dx \, ,
\end{eqnarray}
with $A$ and $B$ constants. This allows us to write a solution for $ \phi $:
\begin{eqnarray}
\phi &=& a^{-\frac{3}{2}}\left( \frac{m}{C} \right)^{-\frac{1}{2}} \left[A\cos{\left(\int\frac{ma^3}{C}d\eta\right)} +B\sin{\left(\int\frac{ma^3}{C}d\eta\right)} \right] \, ,\\
    &=& a^{-\frac{3}{2}}\left( \frac{m}{C} \right)^{-\frac{1}{2}} \left[A\cos{\left(\int m dt\right)} +B\sin{\left(\int m dt\right)} \right] \, ,\\
    \label{phi}
        \phi &=& a^{-\frac{3}{2}} \left[\bar{A}\cos{\left( mt\right)} +\bar{B}\sin{\left( mt\right)} \right] \, .
\end{eqnarray}
Redefining constants as
\begin{eqnarray}\nonumber
   \bar{A} &=& \alpha \sin{\beta} \, ,\\ \nonumber
   \bar{B} &=& \alpha \cos{\beta}\, ,
\end{eqnarray}
using the relation $ \mu = \frac{m}{H_0} $ and defining $ \tau = H_{0} t $, we can rewrite the equation (\ref{phi}) as a function of $N$ and $\tau$:
 \begin{eqnarray}\label{phit}
      \phi &=& \alpha e^{-\frac{3N}{2}} \sin{(\mu \tau+\beta)}\, .
 \end{eqnarray}

Finally, we can write a dimensionless quantity $ \Phi $,
\begin{equation}\label{phiA}
    \Phi \equiv \sqrt{\frac{8 \pi G}{3}}\phi \, ,
\end{equation}
which will give a solution
\begin{eqnarray}\label{Phit}
      \Phi(N) &=& \alpha e^{-\frac{3N}{2}} \sin{(\mu \tau+\beta)}\, .
 \end{eqnarray}
where we have defined $\alpha'\equiv\sqrt{\frac{8 \pi G}{3}}\alpha$ and then we have dropped the prime on $\alpha'$. It allows us to rewrite the dynamic variables $ r$ and $ \theta $ as
\begin{eqnarray}
    r &=& \sqrt{\Omega_{\phi}} = \sqrt{ \frac{1}{H^2} \left[\frac{H_{0}^2}{2} \left(\frac{d\Phi}{d \tau}\right)^2+\frac{m^2 \Phi^2}{2} \right]} \, , \\
    \theta &=& \tan^{-1}{\frac{\mu \Phi}{{d \Phi / d \tau}}}\, ,
\end{eqnarray}
with $\frac{d}{dt}=H_0\frac{d}{d\tau}\,$.

The initial conditions are given in $ \tau = 0 $,
\begin{align}
 \Omega_{\phi0} &= \alpha^2 \left[\frac{9 \sin^2{\beta}}{8}-\frac{3 \mu \sin{2\beta}}{4}+\frac{\mu^2}{2} \right]\, ,\\
 \theta_0 &= \tan^{-1}{\frac{\mu \sin{\beta}}{{\mu \cos{\beta}-\frac{3}{2}\sin{\beta}}}}\,,
 \end{align}
 from which we can write $\alpha$ as a function of $\beta$:
\begin{equation}
    \alpha = \sqrt{\frac{8 \Omega_{\phi 0}}{9\sin^2{\beta}-6\mu \sin{2\beta}+4\mu^2}}\,.
    \label{alfabeta}
\end{equation}

It is important to mention that the denominator in \eqref{alfabeta} is always positive for any $\beta$ and for $\mu\neq0$. Using relation (\ref{phiA}), the Friedmann equation becomes \cite{js}:
\begin{equation}
\left(\frac{H}{H_0}\right)^2=\frac{\frac{1}{2}\mu^2\Phi(N)^2+\Omega_{b0}e^{-3N}+\Omega_{\Lambda0}}{1-\frac{1}{2}\Phi'(N)^2}\,,
\label{eqH2}
\end{equation}
which allows us to find a relation for $ \frac{dN}{d\tau} = \frac{H}{H_0}$:
\be
\frac{dN}{d\tau}=\frac{\sqrt{\Delta}-3\mu\alpha^2\sin\left[2(\beta+\mu\tau)\right]}{8e^{3N}-9\alpha^2\sin^2(\beta+\mu\tau)}
\label{dNdtau}
\ee
where
\be
\Delta\equiv\left[8e^{3N}-9\alpha^2\sin^2(\beta+\mu\tau)\right]\left[8\left(\Omega_b+\Omega_\Lambda e^{3N}\right)+4\mu^2\alpha^2\sin^2(\beta+\mu\tau)\right]+32\mu^2\alpha^2e^{3N}\cos^2(\beta+\mu\tau)
\ee

Defining $\gamma \equiv \frac{1}{\mu}$,
%and using (\ref{phit}), (\ref{phiA}) and the relation above for $dN/d\tau$ in (\ref{eqH2}), 
we can expand \eqref{dNdtau} in series around $\gamma=0$ ($\mu\rightarrow\infty$), obtaining to the first order:
\begin{eqnarray}
  \frac{dN}{d\tau} \simeq   E_{\Lambda} - \frac{3\gamma}{4}e^{-3N}\Omega_{\phi}\left\{\sin\left[2\left(\beta+\frac{\tau}{\gamma}\right)\right] - e^{\frac{3N}{2}}\sin{\left( 2\beta \right)}\right\} \,,
\label{dNdtauap}
\end{eqnarray}
where $E_{\Lambda}=\sqrt{(\Omega_b + \Omega_{\phi})e^{-3N}+\Omega_{\Lambda}}$
and the second term corresponds to a first order correction. From this result we can see that the $\Lambda$CDM background behaviour is a limit for SFDM when $m\rightarrow\infty$. In order to solve it numerically we consider $\tau=\tau_\Lambda$, where $\tau_\Lambda$ is the age in $\Lambda$CDM model. That is, as $\tau$ only appears in the first order term, replacing it by the zeroth order approximation, keeps it correct at first order on $\gamma$.

\section{\label{data}Samples}
\subsection{\texpdf{$H(z)$}  \,\,dataset}
In order to constrain the free parameters, we use the Hubble parameter ($H(z)$) data in different redshift values. These kind of observational data are quite reliable because in general such observational data are independent of the background cosmological model, just relying on astrophysical assumptions. We have used the currently most complete compilation of $H(z)$ data, with 51 measurements \cite{MaganaEtAl18}.

At the present time, the most important methods for obtaining $H(z)$ data are\footnote{See \cite{zt} for a review.} (i) through ``cosmic chronometers'', for example, the differential age of galaxies (DAG) \cite{Simon05,Stern10,Moresco12,Zhang12,Moresco15,MorescoEtAl16}, (ii) measurements of peaks of baryonic acoustic oscillations (BAO) \cite{Gazta09,Blake12,Busca12,AndersonEtAl13,Font-Ribera13,Delubac14} and (iii) through correlation function of luminous red galaxies (LRG) \cite{Chuang13,Oka14}.

Among these methods for estimating $H(z)$, the 51 data compilation as grouped by \cite{MaganaEtAl18}, consists of 20 clustering (BAO+LRG) and 31 differential age $H(z)$ data.

Differently from \cite{MaganaEtAl18}, we choose not to use $H_0$ in our main results here, due to the current tension among $H_0$ values estimated from different observations \citep{RiessEtAl16,Planck16,BernalEtAl16}.

\subsection{SNe Ia}
%Next, we considered a SNe Ia data sample, which, although being more dependent on the fiducial cosmological model, it consists of a large data sample.
% and has passed recently through more refined and model independent methods of light curve fitting \cite{SuzukiEtAl12}.
We have chosen to work with one of the largest SNe Ia sample to date, namely, the Pantheon sample \cite{pantheon}. This sample consists of 279 SNe Ia from Pan-STARRS1 (PS1) Medium Deep Survey ($0.03<z<0.68$), combined with distance estimates of SNe Ia from Sloan Digital Sky Survey (SDSS), SNLS and various low-$z$ and Hubble Space Telescope samples to form the largest combined sample of SNe Ia, consisting of a total of 1048 SNe Ia in the range of $0.01<z<2.3$.% \textcolor{red}{with SNe Ia distance estimates from the Sloan Digital Sky Survey (SDSS), SNLS with low redshift samples $ 0.01 <z <2.3 $ from the Hubble Space Telescope. This compiled data forms the largest SNe Ia database with 1048 values in the redshift range of $ 0.01 <z <2.3 $.}

As explained on \cite{pantheon}, the PS1 light-curve fitting has been made with SALT2 \cite{GuyEtAl10}, as it has been trained on the JLA sample \cite{BetouleEtAl14}. %\textcolor{red}{as it was done in the sample JLA  \cite{BetouleEtAl14}}.
Three quantities are determined in the light-curve fit that are needed to derive a distance: the colour $c$, the light-curve shape parameter $x_1$ and the log of the overall flux normalization $m_B$.%\textcolor{red}{The three parameters are estimated by adjusting the light curve, being essential to derive the distance: the colour $c$ parameter, the light-curve profile $x_1$ and the logarithm of the general flux normalization $m_B$.}

The SALT2 light-curve fit parameters are transformed into distances using a modified version of the Tripp formula \cite{Tripp98},%\textcolor{red}{The light curve parameter estimates are obtained from the SALT2 adjustment, so it is possible to calculate the distances using a modified version of the Tripp formula \cite{Tripp98},}
%we may assume that supernovae with identical color, shape and galactic environment have on average the same intrinsic luminosity for all redshifts. In this case, the distance modulus $\mu=5\log_{10}(d_L(\mathrm{pc})/10)$ may be given as
\begin{equation}
 %\mu=m_B^*-(M_B-\alpha\times X_1+\beta\times C)
 \mu=m_B-M+\alpha x_1-\beta c+\Delta_M+\Delta_B,
 \label{mupanth}
\end{equation}
where $\mu$ is the distance modulus, $\Delta_M$ is a distance correction term based on the host galaxy mass of the SN, and $\Delta_B$ is a distance correction factor based on predicted biases from simulations. As can be seen, $\alpha$ is the coefficient of the relation between luminosity and stretch, while $\beta$ is the coefficient of the relation between luminosity and color, and $M$ is the absolute $B$-band magnitude of a fiducial SN Ia with $x_1=0$ and $c=0$.

Differently from previous SNe Ia samples, like JLA \cite{BetouleEtAl14}, Pantheon uses a calibration method named BEAMS with Bias Corrections (BBC), which allows to determine SNe Ia distances without one having to fit SNe parameters jointly with cosmological parameters. Thus, Pantheon provide directly corrected $m_B$ estimates in order to constrain cosmological parameters alone.

The systematic uncertainties were propagated through a systematic uncertainty matrix. An
uncertainty matrix C was defined such that
%\textcolor{red}{The error calculations were propagated through the systematic uncertainty matrix C which can be defined by}
\be\label{cov}
\mathbf{C} = \mathbf{D}_{\mathrm{stat}} + \mathbf{C}_{\mathrm{sys}}.
\ee

The statistical matrix $\mathbf{D}_\mathrm{stat}$ has only diagonal components that includes photometric errors of the SN distance, the distance uncertainty from the mass step correction, the uncertainty from the distance bias correction, the uncertainty from the peculiar velocity uncertainty and redshift measurement uncertainty in quadrature, the uncertainty from stochastic gravitational lensing, and the intrinsic scatter.

\section{\label{analysis}Analyses and Results}
In our analyses, we have chosen flat priors for all parameters, so always the posterior distributions are proportional to the likelihoods.

For $H(z)$ data, the likelihood distribution function is given by $\like_H \propto e^{-\frac{\chi^2_H}{2}}$, where 
%\begin{eqnarray}
%\chi^2_{SN} &=& (\hat{\bm{\mu}}-{\bm\mu}(z,\theta_{SN},z_t,\theta_{mod,j}))^TC^{-1}(\hat{\bm\mu}-{\bm\mu}(z,\theta_{SN},z_t,\theta_{mod,j}))\\
%\chi^2_H    &=& \sum_{i = 1}^{41}\frac{{\left[ H_{obs,i} - H(z_i,H_0,z_t,\theta_{mod,j})\right] }^{2}}{\sigma^{2}_{H_i,obs}} ,
%\label{chi}
%\end{eqnarray} 
\begin{equation}
\chi^2_H = \sum_{i = 1}^{51}\frac{{\left[ H_{obs,i} - H(z_i,\mathbf{s})\right] }^{2}}{\sigma^{2}_{H_i,obs}} ,
\label{chi2H}
\end{equation}
%where $s_j$ is the specific parameter for each model, it can be $d_2$, $h_2$ or $q_1$, for $D_C(z)$, $H(z)$, $q(z)$ parametrizations, respectively. 

The $\chi^2$ function for Pantheon is given by
\be
\chi^2=\mathbf{\Delta m}^T\cdot\mathbf{C}^{-1}\cdot\mathbf{\Delta m},
\ee
where $\textbf{C}$ is the same from \eqref{cov}, $\mathbf{\Delta m}=m_B-m_{\mathrm{mod}}$, and
\be
m_{\mathrm{mod}}=5\log_{10}D_L(z)+\mathcal{M},
\ee
where $\mathcal{M}$ is a nuisance parameter which encompasses $H_0$ and $M$. We choose to project over $\mathcal{M}$, which is equivalent to marginalize the likelihood $\like\propto e^{-\chi^2/2}$ over $\mathcal{M}$, up to a normalization constant. In this case we find the projected $\chi^2_{proj}$:
\be
\chi^2_{proj}=S_{mm}-\frac{S_m^2}{S_A}
\ee
where $S_{mm}=\sum_{i,j}\Delta m_i\Delta m_jA_{ij}=\mathbf{\Delta m}^T\cdot\mathbf{A}\cdot\mathbf{\Delta m}$, $S_m=\sum_{i,j}\Delta m_iA_{ij}=\mathbf{\Delta m}^T\cdot\mathbf{A}\cdot\mathbf{1}$, $S_A=\sum_{i,j}A_{ij}=\mathbf{1}^T\cdot\mathbf{A}\cdot\mathbf{1}$ and $\mathbf{A}\equiv\mathbf{C}^{-1}$.

In order to obtain the constraints over the free parameters, we have sampled the likelihood $\like\propto e^{-\chi^2/2}$ through Monte Carlo Markov Chain (MCMC) analysis. A simple and powerful MCMC method is the so called Affine Invariant MCMC Ensemble Sampler by \cite{GoodWeare}, which was implemented in {\sffamily Python} language with the {\sffamily emcee} software by \cite{ForemanMackey13}. This MCMC method has the advantage over simple Metropolis-Hastings (MH) methods of depending on only one scale parameter of the proposal distribution and on the number of walkers, while MH methods in general depend on the parameter covariance matrix, that is, it depends on $n(n+1)/2$ tuning parameters, where $n$ is dimension of parameter space. The main idea of the Goodman-Weare affine-invariant sampler is the so called ``stretch move'', where the position (parameter vector in parameter space) of a walker (chain) is determined by the position of the other walkers. Foreman-Mackey {\it et al.} modified this method, in order to make it suitable for parallelization, by splitting the walkers in two groups, then the position of a walker in one group is determined by {\it only} the position of walkers of the other group\footnote{See \cite{AllisonDunkley14} for a comparison among various MCMC sampling techniques.}.

\begin{figure}[h!]
 \centering
 \includegraphics[width=.8\textwidth]{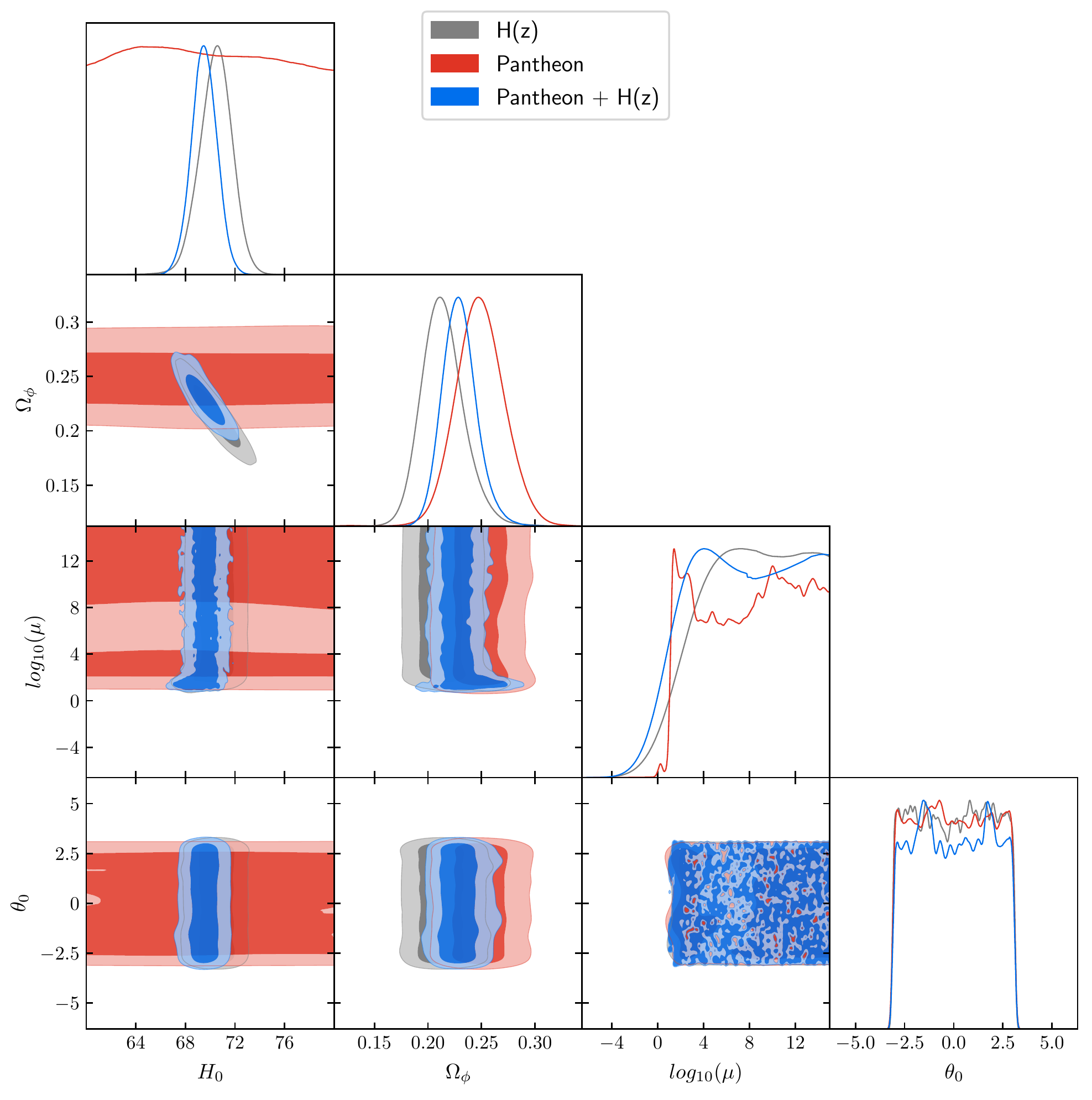}
 %\hspace{.1\textwidth}
 %\includegraphics[width=.3\textwidth]{fig/RiessCLs.png}%{fig/SCP98.png}
 % HrmWoE.png: 801x1064 px, 72dpi, 28.26x37.54 cm, bb=0 0 801 1064
 \caption{Constraints from Pantheon SNe Ia and $H(z)$.}
 \label{DcPolinTriangAll}
\end{figure}

\begin{figure}[h!]
 \centering
 \includegraphics[width=.8\textwidth]{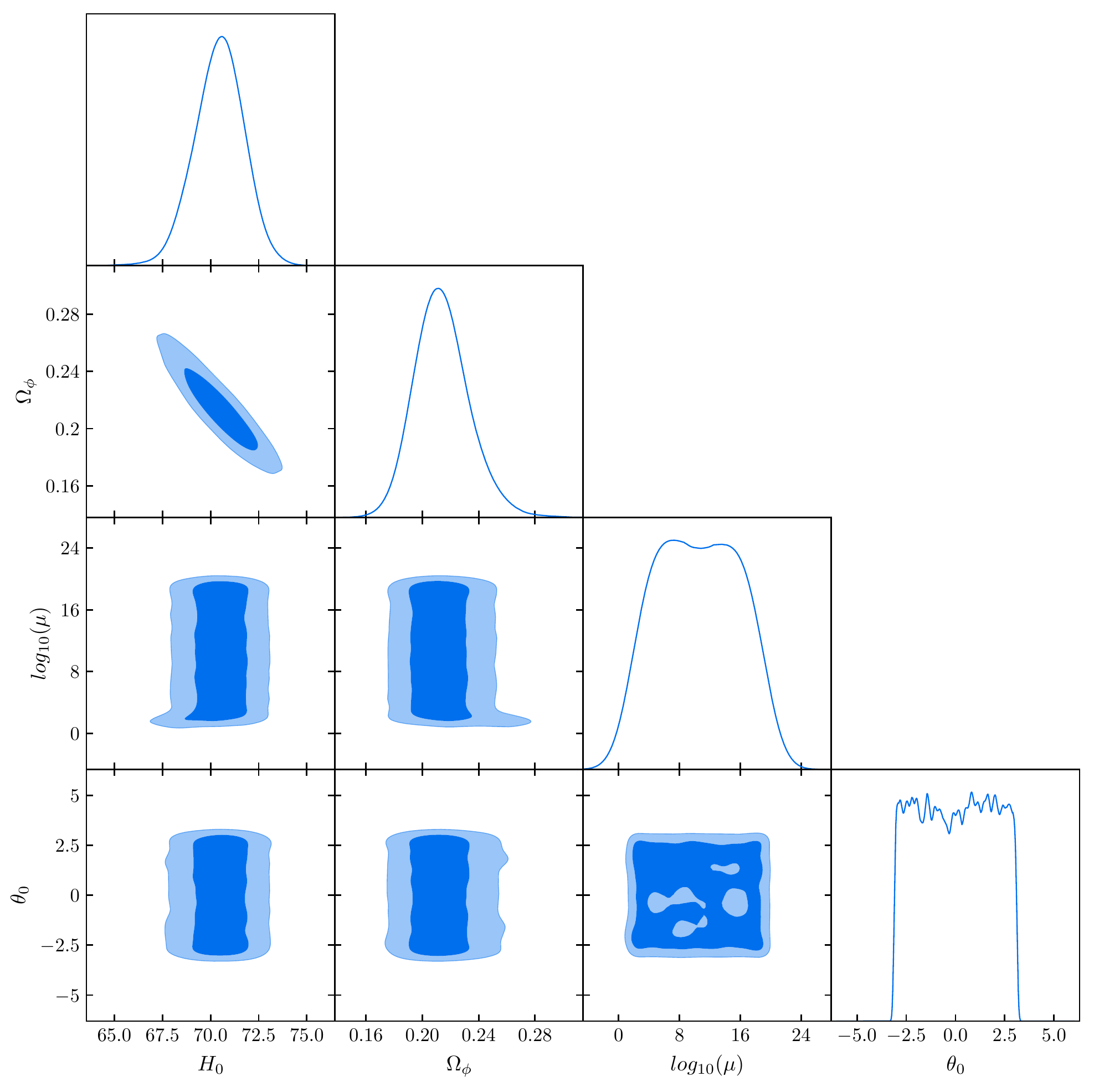}
 %\hspace{.1\textwidth}
 %\includegraphics[width=.3\textwidth]{fig/RiessCLs.png}%{fig/SCP98.png}
 % HrmWoE.png: 801x1064 px, 72dpi, 28.26x37.54 cm, bb=0 0 801 1064
 \caption{Constraints from Pantheon SNe Ia and $H(z)$.}
 \label{ztlikes}
\end{figure}

We used the freely available software {\sffamily emcee} to sample from our likelihood in $n$-dimensional parameter space. We have used flat priors over the parameters.
In order to plot all the constraints on each model in the same figure, we have used the freely available software {\sffamily getdist}\footnote{{\sffamily getdist} is part of the great MCMC sampler and CMB power spectrum solver {\sffamily COSMOMC}, by \cite{cosmomc}.}, in its {\sffamily Python} version. The results of our statistical analyses can be seen on Figs. \ref{DcPolinTriangAll}-\ref{ztlikes} and on Table \ref{tab1}.

The lower limit in this context, with  $2\sigma$ c.l., is $\mu=0.52$. It corresponds to a dark matter mass $m=0.52H_0$, that is, $m= 5.1\times10^{-34}\text{ eV}$ for a Hubble constant of $H_0=69.5^{+2.0}_{-2.1}\text{ km s}^{-1}\text{Mpc}^{-1}$, found from the statistical analysis.

%It must be noted, also, that Maga\~na and Matos have not made an statistical analysis of SFDM to find the DM mass estimate. Instead, they have preferred to make use of control parameter $k$ in order to have dynamical evolution similar to $\Lambda$CDM.

%\begin{figure}[t]
%\centerline{\epsfig{figure=sfdmHzData.eps,width=0.70\linewidth,angle=0}}%-90}}height=3.8truein,
%\caption{$H(z)$ data from Sharov and Vorontsova (2014) \cite{sharov} and some $H(z)$ curves with $h$ fixed on the best fit ($h=0.673$) and $\mu_{100}$ varying from the best fit, to 1$\sigma$ below and 1$\sigma$ above the best fit.}
%\label{sfdmEvo}
%\end{figure}

\begin{table}
\centering
\begin{tabular} { l  c}
\hline
 Parameter &  95\% limits\\
\hline
{\boldmath$H_0            $} & $69.5^{+2.0}_{-2.1}        $\\

{\boldmath$\Omega_\phi    $} & $0.230^{+0.033}_{-0.031}   $\\

{\boldmath$log_{10}(\mu)  $} & $>\,\, -1.28          $\\

{\boldmath$\theta_0       $} & $0.0^{+3.0}_{-3.0}         $\\
\hline
\end{tabular}
\caption{Mean Parameter Values  }
\label{tab1}
\end{table}
%%%%%%%%%%%%%%%%%%%%%%%%%%%%%%%%%%%%%%%%%%%%%%%%%%%%%%%%%%%%%%%%%%%%%%%%%%
\section{Concluding remarks\label{conclusion}}
We have studied the SFDM model, which hypothesizes the description of dark matter by a real scalar field $ \phi $. Taking a potential of the free field type, we have performed an statistical analysis of the model using tools such as the JWKB approximation, which allowed the numerical integrations of the \eqref{eq:pol_r}--\eqref{eq:pol_q} system to be executed for high values of $ \mu $ faster than the exact solution. In the statistical analysis, the use of the MCMC method guaranteed greater precision in the results of the analysis of the $ 51 $ data of $ H (z) $ and the Pantheon sample of the SNe Ia data.

We have established limits for the free parameters of the model, obtaining a lower limit for the mass of the dark matter particle of about $10^{-34}\text{ eV}$ with $2 \sigma $ confidence, a value close to what was found in \cite{js, Marsh} ($\approx10^{-33}$ eV). In our analysis, we have also obtained $H_0=69.5^{+2.0}_{-2.1} $ km s$^{-1}$Mpc$^{-1}$. The value of $ \Omega_{\phi} = 0.230^{+0.033}_{-0.031} $  found and, together with the well-known $ \Omega_b = 0.032^{+0.092}_{-0.097} $, corresponding to baryonic matter in the universe, provides a total of $  \Omega_M = 0.262^{+0.098}_{-0.102}$. Both results for $H_0$ and $ \Omega_M$ are compatible with the result found by Planck, of $H_0=67.4 \pm 0.5 $ km s$^{-1}$Mpc$^{-1}$ and $  \Omega_M = 0.315\pm 0.007$. %colocar os valores de H0 e Omega_m do Planck

The SFDM model was satisfactory in describing astronomical observations reproducing results compatible with $\Lambda$CDM model and yet, when the mass of dark matter is very high, the SFDM model tends to $\Lambda$CDM. The advantage of the alternative model is in the explanation of some small scale CDM problems \cite{pp}, like the cuspy/core problem. Since the dark matter particles form Bose-Einstein condensates very early in the evolution of the universe due to their low mass, it leads to the creation of a flat density profile in the center of galaxies \cite{2}.

%Experiments involving detection of dark matter particles, for example, DAMA \cite{dama}, search for particles with mass in the range $( \approx 15 - 120 GeV)$. More recently, in \cite{bene}, an upper limit for the dark matter mass of $ \approx $ 197 TeV has been established, based on the abundance of relics of thermal dark matter particles annihilated through long-range interaction.

We have seen that the real scalar field $ \phi $ proved to be a promising candidate for dark matter, although further analysis and testing should be done with the model to refine its results, for instance, by using CMB data that carries information from the primordial Universe. 

\begin{acknowledgments}
This study was financed in part by the Coordenação de Aperfeiçoamento de Pessoal de Nível Superior - Brasil (CAPES) - Finance Code 001. JFJ is supported by Funda\c{c}\~ao de Amparo \`a Pesquisa do Estado de S\~ao Paulo - FAPESP (Processes no. 2013/26258-4 and 2017/05859-0). SHP acknowledges financial support from  {Conselho Nacional de Desenvolvimento Cient\'ifico e Tecnol\'ogico} (CNPq)  (No. 303583/2018-5 and 400924/2016-1).
\end{acknowledgments}

\end{document}